\documentclass[twocolumn,floatfix,aps,eqsecnum,prb]{revtex4-2}
\usepackage{graphicx}
\usepackage{amsmath}
\usepackage{amssymb}
\usepackage{bm}

\usepackage{color}

\begin{document}

\title{Voltage staircase in a current-biased quantum-dot Josephson junction}
\author{D. O. Oriekhov}
\affiliation{Instituut-Lorentz, Universiteit Leiden, P.O. Box 9506, 2300 RA Leiden, The Netherlands}
\author{Y. Cheipesh}
\affiliation{Instituut-Lorentz, Universiteit Leiden, P.O. Box 9506, 2300 RA Leiden, The Netherlands}
\author{C. W. J. Beenakker}
\affiliation{Instituut-Lorentz, Universiteit Leiden, P.O. Box 9506, 2300 RA Leiden, The Netherlands}
\date{February 2021}
\begin{abstract}
We calculate the current-voltage ($I$-$V$) characteristic of a Josephson junction containing a resonant level in the weakly coupled regime (resonance width small compared to the superconducting gap). The phase $\phi$ across the junction becomes time dependent in response to a {\sc dc} current bias. Rabi oscillations in the Andreev levels produce a staircase $I$-$V$ characteristic. The number of voltage steps counts the number of Rabi oscillations per $2\pi$ increment of $\phi$, providing a way to probe the coherence of the qubit in the absence of any external {\sc ac} driving. The phenomenology is the same as the \textit{``Majorana-induced DC Shapiro steps in topological Josephson junctions''} of  Phys.\ Rev.\ B \textbf{102}, 140501(R) (2020) --- but now for a non-topological Andreev qubit.
\end{abstract}
\maketitle

\section{Introduction}
\label{intro}

A single-mode weak link between superconductors supports a two-level system with a spacing that is adjustable via the superconducting phase difference \cite{Fur91,Bee91a}. Because Andreev reflection is at the origin of the phase sensitivity, the levels are called Andreev levels. Although their existence was implicit in early studies of the Josephson effect \cite{Kul77}, the characteristic dependence $\propto\sqrt{1-\tau\sin^2(\phi/2)}$ of the level spacing on the phase $\phi$, with $\tau$ the transmission probability, was only identified \cite{Bee91b} with the advent of nanostructures. The present interest in quantum information processing is driving theoretical \cite{Zaz03,Cht03} and experimental \cite{Bre13,Jan15,Hay18,Hay21} studies of Andreev levels as qubits. 

To assess the coherence of the qubit one would use \textsc{ac} microwave radiation of the two-level system and perform a time-resolved detection of the Rabi oscillations of the wave function \cite{Ber11}. In this work we will show how a \textsc{dc} current $I_{\textsc{dc}}$ and measurement of the time-averaged voltage $\bar{V}$ can be used to detect Rabi oscillations of an Andreev qubit: The staircase dependence of $\bar{V}$ on $I_{\textsc{dc}}$ counts the number of Rabi oscillations per $2\pi$ increment of $\phi$.

Our study is motivated by Choi, Calzona, and Trauzettel's report \cite{Cho20} of such a remarkable effect (dubbed ``\textsc{dc} Shapiro steps'') in a Majorana qubit --- which is the building block of a topological quantum computer. As we will see, neither the unique topological properties of a Majorana qubit (its non-Abelian braiding and fusion rules) nor its specific symmetry class (class D, with broken time-reversal and spin-rotation symmetry) are needed,  a similar phenomenology can be found in a non-topological Andreev qubit with preserved symmetries (class CI).

The outline of this paper is as follows. In the next section \ref{sec_AlevelH} we present the model of the weak link that we will consider: a quantum dot connecting two superconductors with a tunnel rate $\Gamma$ small compared to the superconducting gap $\Delta_0$. Such a Josephson junction has been extensively studied \cite{Mar11,Med19,Lev03} in the regime where Coulomb charging and the Kondo effect govern the charge transfer \cite{Gla89,Spi91,Nov05}. We will assume the charging energy is small and treat the quasiparticles as noninteracting. 

The dynamics of a current-biased, resistively shunted quantum-dot Josephson junction is studied in Secs.\ \ref{sec_voltage} and \ref{sec_dynamics}. The voltage staircase is shown in Fig.\ \ref{fig_IV} and the one-to-one relationship with the number of Rabi oscillations is in Fig.\ \ref{fig_counter}. In the concluding section \ref{sec_discuss} we will explain why the substitution of the quantum dot by a quantum point contact will remove the voltage staircase.

\section{Andreev level Hamiltonian}
\label{sec_AlevelH}

\begin{figure}[tb]
\centerline{\includegraphics[width=1\linewidth]{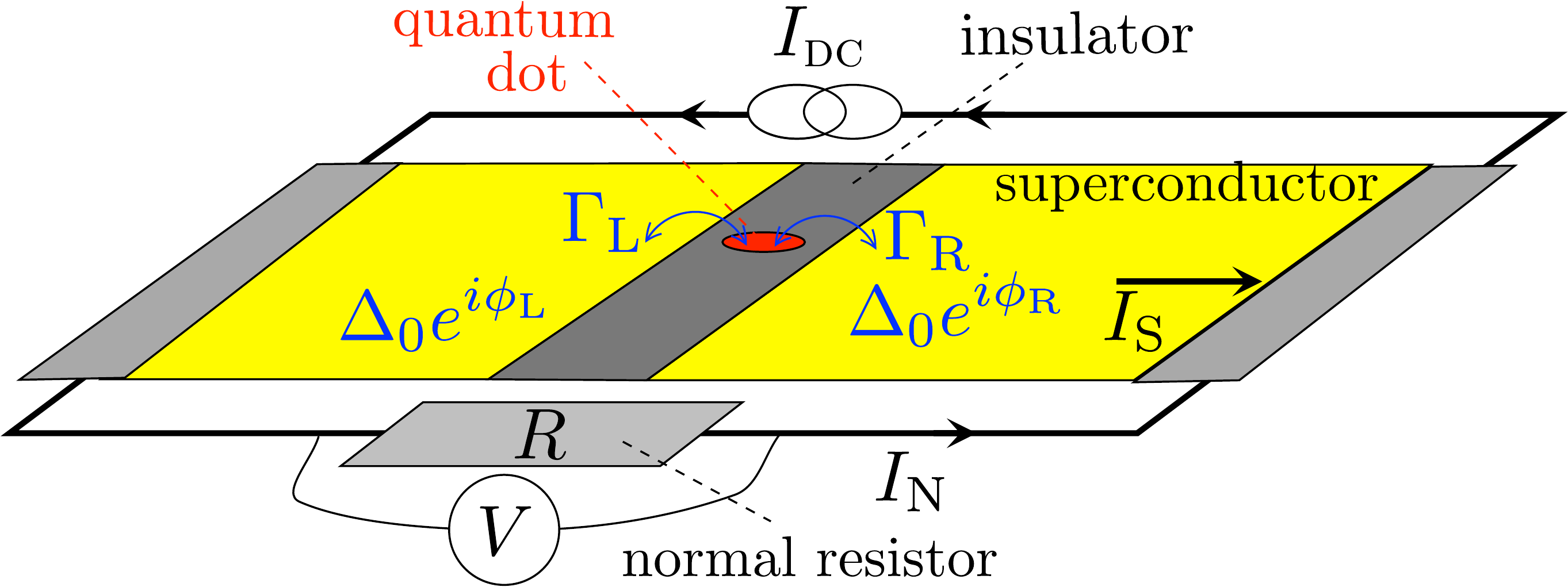}}
\caption{Current-biased, resistively-shunted Josephson junction, formed out of two superconductors (phases $\phi_{\rm L}$ and $\phi_{\rm R}$) separated by an insulator containing a quantum dot (tunnel rates $\Gamma_{\rm L}$ and $\Gamma_{\rm R}$ from the left and from the right). The superconducting phases become time dependent when a voltage difference $V$ develops in response to a {\sc dc} current $I_{\textsc{dc}}$.
}
\label{fig_setupQDJJ}
\end{figure}

We consider the Josephson junction shown in Fig.\ \ref{fig_setupQDJJ}, consisting of a quantum dot in the normal state (N) coupled via a tunnel barrier to superconductors (S) at the left and right, with pair potentials $\Delta_0 e^{i \phi_{\rm L}}$ and $\Delta_0 e^{i \phi_{\rm R}}$. We focus on the weakly coupled regime, when the tunnel rates $\Gamma_{\rm L}$ and $\Gamma_{\rm R}$ through the barrier are small compared to $\Delta_0$. 

We assume that the fully isolated quantum dot has a single electronic energy level $E_0$ within an energy range $\Gamma=\Gamma_{\rm L}+\Gamma_{\rm R}$ from the Fermi energy $\mu$. The normal-state conductance $G_{\rm N}$ is then given by the Breit-Wigner formula
\begin{equation}
G_{\rm N}=\frac{2e^2}{h}\tau_{\rm BW},\;\;\tau_{\rm BW}=\frac{\Gamma_{\rm L}\Gamma_{\rm R}}{(E_0-\mu)^2+\tfrac{1}{4}\Gamma^2}.\label{GBW}
\end{equation}
Coupling of electrons and holes by Andreev reflection from the superconductor produces a pair of Andreev levels at energies $\pm E_{\rm A}(\phi)$, dependent on the phase difference $\phi=\phi_{\rm L}-\phi_{\rm R}$ between the left and right superconductors.

A simplifying assumption of our analysis is that the Coulomb charging energy $U$ is small compared to $\Gamma$ and can be neglected. If $U$ is larger than $\Gamma$ but still smaller than $\Delta_0$, the main effect of the charging energy is a shift of the energy level of the dot, $E_0\mapsto E_0+U/2$. Provided $E_0>0$ the ground state remains a spin-singlet \cite{Men09}, and we do not expect a qualitative change in our results. If $U$ becomes larger than $\Delta_0$ the supercurrent is reduced by a factor $\Gamma/\Delta_0$ because tunneling of a Cooper pair into the quantum dot is suppressed \cite{Gla89,Spi91,Nov05}.

To describe the non-equilibrium dynamics of the junction we seek the effective low-energy Hamiltonian of time-dependent Andreev levels. This requires information not only on the eigenvalues but also on the eigenfunctions. In subsections \ref{sec_Andreevlevel} and \ref{sec_effHdc} we summarize results from Refs.\ \onlinecite{Men09,Bee92,Dev97,Rec10} for the time-independent situation, which we need as input for the dynamical study starting from subsection \ref{sec_effHac}.

\subsection{Andreev levels}
\label{sec_Andreevlevel}

\begin{figure}[tb]
\centerline{\includegraphics[width=0.7\linewidth]{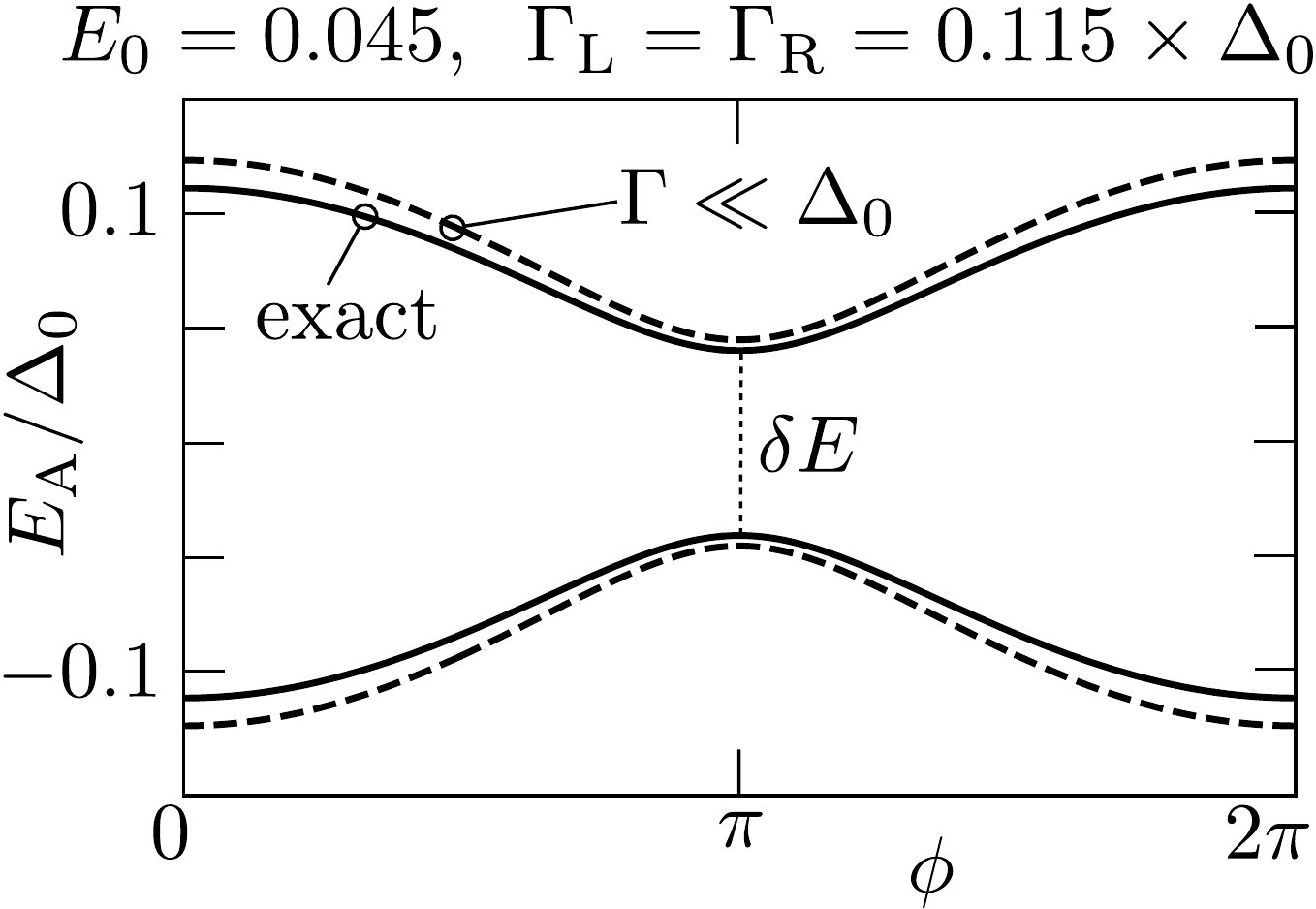}}
\caption{Andreev levels $\pm E_{\rm A}(\phi)$ according to the full expression \eqref{Omegaeq} (solid curve) and in the weak-coupling approximation \eqref{EAweakcoupling} (dashed curve, parameters $E_0=0.045$, $\mu=0$, $\Gamma_{\rm L}=\Gamma_{\rm R}=0.115$, all in units of $\Delta_0$).
}
\label{fig_Evsphi}
\end{figure}

For arbitrary ratio of $\Gamma$ and $\Delta_0$ the energies of the Andreev levels are equal to the two real solutions $\pm E_{\rm A}$ of the equation \cite{Bee92,Dev97}
\begin{equation}
\Omega(E,\phi)+\Gamma E^2\sqrt{\Delta_0^2-E^2}=0,\label{Omegaeq}
\end{equation}
with
\begin{align}
\Omega(E,\phi)={}&(\Delta_0^2-E^2)\bigl[E^2-(E_0-\mu)^2-\tfrac{1}{4}\Gamma^2\bigr]\nonumber\\
&+\Delta_0^2\Gamma_{\rm L}\Gamma_{\rm R}\sin^2(\phi/2).
\end{align}
In the weak-coupling regime $\Gamma\ll\Delta_0$, assuming also $|E_0-\mu|\ll\Delta_0$, this reduces to
\begin{equation}
\begin{split}
&E_{\rm A}=\Delta_{\rm eff}\sqrt{1-\tau_{\rm BW}\sin^2(\phi/2)},\\
&\Delta_{\rm eff}=\sqrt{(E_0-\mu)^2+\tfrac{1}{4}\Gamma^2},
\end{split}
\label{EAweakcoupling}
\end{equation}
no longer dependent on $\Delta_0$. The two Andreev levels have an avoided crossing at $\phi=\pi$, separated by an energy
\begin{equation}
\delta E=\sqrt{4(E_0-\mu)^2 +(\Gamma_{\rm L}-\Gamma_{\rm R})^2},
\end{equation}
see Fig.\ \ref{fig_Evsphi}.

The equilibrium supercurrent, at temperatures $k_{\rm B}T\ll\Gamma$, is given by
\begin{equation}
I_{\rm eq}(\phi)=-\frac{2e}{\hbar}\frac{dE_{\rm A}}{d\phi}=\frac{e\Gamma_{\rm L}\Gamma_{\rm R}\sin\phi}{2\hbar E_{\rm A}(\phi)},\label{Ieqphi}
\end{equation}
with critical current (maximal supercurrent)
\begin{align}
I_{\rm c}={}&\frac{e}{\hbar}\biggl(\sqrt{(E_0-\mu)^2+\tfrac{1}{4}\Gamma^2}\nonumber\\
& -\sqrt{(E_0-\mu)^2+\tfrac{1}{4}\Gamma^2-\Gamma_{\rm L}\Gamma_{\rm R}}\biggr).\label{Icresult}
\end{align}
There is no contribution from the continuous spectrum in the weak-coupling regime \cite{Bee92}.

\subsection{Effective Hamiltonian: time-independent phase}
\label{sec_effHdc}

For time-independent phases the effective low-energy Hamiltonian in the weak-coupling regime $\Gamma\ll \Delta_0$ follows from second-order perturbation theory \cite{Men09,Rec10},
\begin{align}
H={}&-\tfrac{1}{2}\bigl(e^{i\phi_{\rm L}}\Gamma_{\rm L}+e^{i\phi_{\rm R}}\Gamma_{\rm R}\bigr)a^\dagger_\uparrow a^\dagger_\downarrow+\text{H.c.} \nonumber\\
&+(E_0-\mu)(a^\dagger_\uparrow a_\uparrow^{\vphantom{\dagger}}+a^\dagger_\downarrow a_\downarrow^{\vphantom{\dagger}}).
\end{align}
Here $a_\uparrow$ and $a_\downarrow$ are the fermionic annihilation operators of a spin-up or spin-down electron in the quantum dot.

The corresponding Bogoliubov-De Gennes (BdG) Hamiltonian ${\cal H}$ is a $4\times 4$ matrix contracted with the spinors $\Psi=(a_\uparrow,-a_\downarrow^\dagger,a_\downarrow,-a_\uparrow^\dagger)$ and $\Psi^\dagger$,
\begin{equation}
H=\tfrac{1}{2}\Psi^\dagger\cdot{\cal H}\cdot\Psi+E_0-\mu.
\end{equation}
It is block-diagonal, so we only need to consider one $2\times 2$ block, given by
\begin{equation}
{\cal H}=\begin{pmatrix}
E_0-\mu&\tfrac{1}{2}e^{i\phi_{\rm L}}\Gamma_{\rm L}+\tfrac{1}{2}e^{i\phi_{\rm R}}\Gamma_{\rm R}\\
\tfrac{1}{2}e^{-i\phi_{\rm L}}\Gamma_{\rm L}+\tfrac{1}{2}e^{-i\phi_{\rm R}}\Gamma_{\rm R}&\mu-E_0
\end{pmatrix}.\label{Htimeindependent}
\end{equation} 
One readily checks that the eigenvalues $\pm E_{\rm A}$ of ${\cal H}$ are given by Eq.\ \eqref{EAweakcoupling}.

\subsection{Effective Hamiltonian: time-dependent phase}
\label{sec_effHac}

When the left and right superconductors are at different voltages $\pm V/2$, the superconducting phase becomes time dependent. We choose a gauge such that $\phi_{\rm L}(t)=\phi(t)/2$, $\phi_{\rm R}(t)=-\phi(t)/2$, evolving in time according to the Josephson relation
\begin{equation}
\dot{\phi}\equiv d\phi/dt=(2e/\hbar)V.\label{JJrelation} 
\end{equation}
The voltage bias imposes an electrical potential on the quantum dot, which shifts $\mu$ by an amount $\tfrac{1}{2}\gamma eV$ with $\gamma=(\Gamma_{\rm L}-\Gamma_{\rm R})/\Gamma$.
\begin{widetext}
The time dependent BdG Hamiltonian then becomes
\begin{align}
{\cal H}(t)&=\begin{pmatrix}
E_0-\mu-\tfrac{1}{4}\hbar\gamma \dot{\phi}(t)&\tfrac{1}{2}e^{i\phi(t)/2}\Gamma_{\rm L}+\tfrac{1}{2}e^{-i\phi(t)/2}\Gamma_{\rm R}\\
\tfrac{1}{2}e^{-i\phi(t)/2}\Gamma_{\rm L}+\tfrac{1}{2}e^{i\phi(t)/2}\Gamma_{\rm R}&\mu-E_0+\tfrac{1}{4}\hbar\gamma \dot{\phi}(t)
\end{pmatrix}\nonumber\\
&=\bigl[E_0-\mu-\tfrac{1}{4}\hbar\gamma \dot{\phi}(t)\bigr]\sigma_z+\tfrac{1}{2}\Gamma\bigl[\sigma_x\cos\tfrac{1}{2}\phi(t)-\gamma\sigma_y\sin\tfrac{1}{2}\phi(t)\bigr].\label{Htdef}
\end{align}
\end{widetext}
The Pauli matrices act on the electron-hole degree of freedom. The corresponding current operator is given by
\begin{equation}
{I}(t)=\frac{2e}{\hbar}\frac{\partial}{\partial \phi}{\cal H}(t)=-\frac{e\Gamma}{2\hbar}\bigl[\sigma_x\sin\tfrac{1}{2}\phi(t)+\gamma\sigma_y\cos\tfrac{1}{2}\phi(t)\bigr].\label{Itdef}
\end{equation}

Notice that the Hamiltonian \eqref{Htdef} depends both on $\phi(t)$ and on $\dot{\phi}(t)$, unless $\Gamma_{\rm L}=\Gamma_{\rm R}$. It is possible to remove the $\dot{\phi}$-dependence by a time-dependent unitary transformation \cite{note1}, but since this does not simplify our subsequent calculations we will keep the form \eqref{Htdef}.

\section{Voltage staircase}
\label{sec_voltage}

As shown in Fig.\ \ref{fig_setupQDJJ}, a time-independent current bias $I_{\textsc{dc}}$ is driven partially through the Josephson junction, as a supercurrent $I_{\rm S}(t)$, and partially through a parallel resistor $R$ as a normal current $I_{\rm N}(t)=V(t)/R$. Substitution of the Josephson relation \eqref{JJrelation} gives the differential equation
\begin{equation}
d\phi(t)/dt=(2eR/\hbar)[I_{\textsc{dc}}-I_{\rm S}(t)].
\end{equation}
Here we neglect the junction capacitance (overdamped regime of a resistively shunted Josephson junction) \cite{tinkham}. We work in the low-temperature regime, $k_{\rm B}T\ll\Delta_0$, so that we may ignore thermal fluctuations of the phase due to the voltage noise over the external resistance \cite{Ave98}. 

The supercurrent is obtained from the expectation value
\begin{equation}
I_{\rm S}(t)=\langle\Psi(t)|I(t)|\Psi(t)\rangle,
\end{equation}
where the current operator is given by Eq.\ \eqref{Itdef} and the wave function evolves according to the Schr\"{o}dinger equation
\begin{equation}
i\hbar\frac{d}{dt}|\Psi(t)\rangle={\cal H}(t)|\Psi(t)\rangle.\label{Schreq}
\end{equation}

As initial condition we take $\phi(0)=0$ and $|\Psi(0)\rangle$ the eigenstate of the Andreev level at $-E_{\rm A}$ for $\phi=0$. The {\sc dc} current $I_{\textsc{dc}}$ is increased slowly from zero to some maximal value and then slowly decreased back to zero. The $I$--$V$ characteristic is obtained by averaging $V(t)$ over a moving time window in which $I_{\textsc{dc}}$ is approximately constant. 

\begin{figure}[tb]
\centerline{\includegraphics[width=1\linewidth]{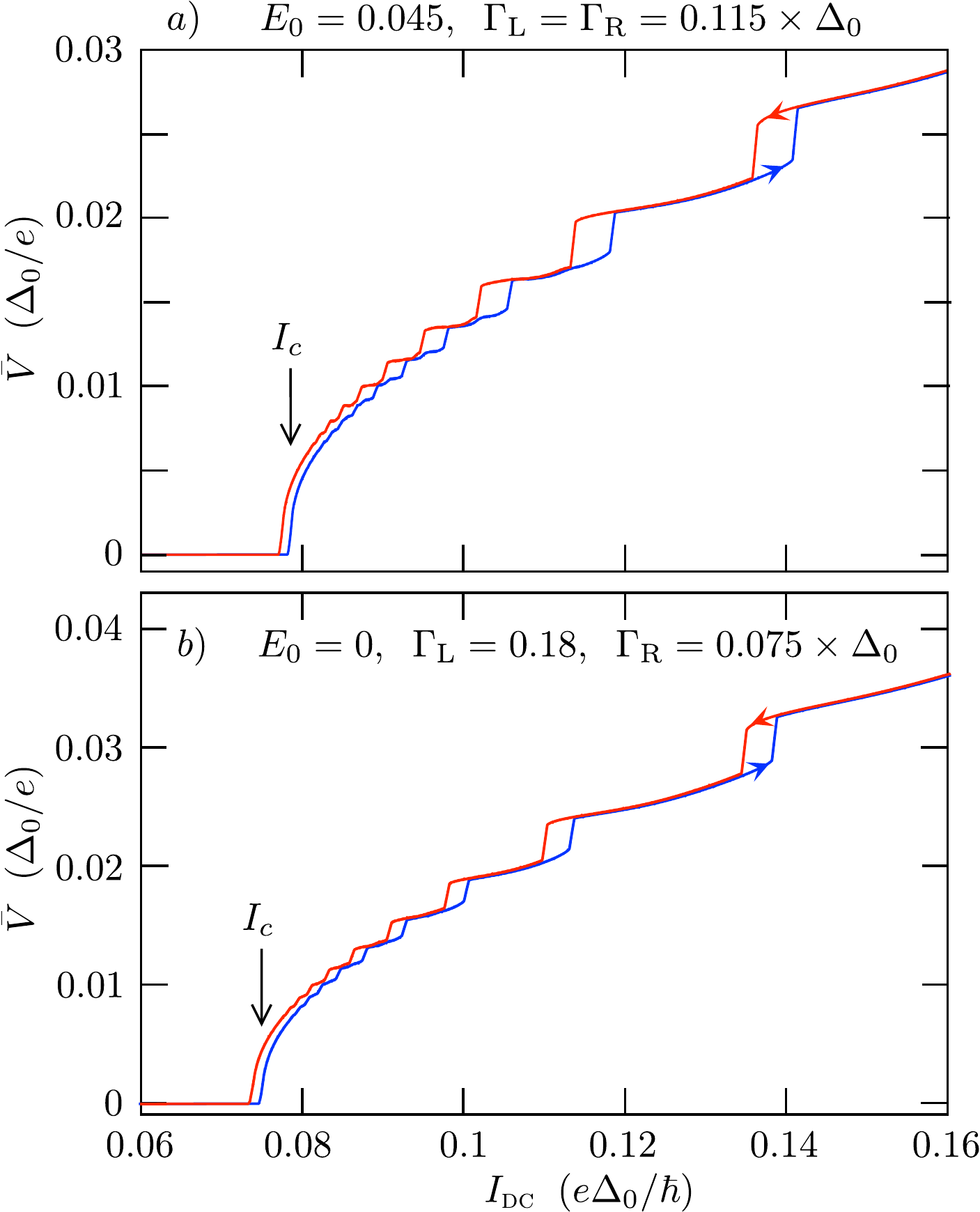}}
\caption{Current-voltage characteristic of the quantum-dot Josephson junction, for two different parameter sets \cite{note2}. The blue curve is for increasing {\sc dc} current, the red curve for decreasing current. The Andreev levels in Fig.\ \ref{fig_Evsphi} correspond to the parameters in panel a). The critical current \eqref{Icresult} is indicated by the black arrow.
}
\label{fig_IV}
\end{figure}

Results of this numerical integration are shown in Fig.\ \ref{fig_IV}. We observe a staircase dependence of $\bar{V}$ on $I_{\textsc{dc}}$. The nonzero voltage appears at the critical current \eqref{Icresult} for the up-sweep and disappears at a slightly lower current for the down sweep. (A similar difference between switching current and retrapping current was found for the Majorana qubit \cite{Fen18}.) The voltage steps at $I_{\textsc{dc}}>I_c$ also show hysteresis: the voltage jump up happens at larger {\sc dc} current than the voltage jump down. (This hysteresis also appears in the Majorana qubit, see App.\ \ref{app_hysteresis}.)

\section{Andreev qubit dynamics}
\label{sec_dynamics}

The voltage staircase of Fig.\ \ref{fig_IV} is a signature of Rabi oscillations of the Andreev qubit formed by the two Andreev levels in the Josephson junction, in much the same way that the voltage steps of Ref.\ \onlinecite{Cho20} were driven by Rabi oscillations of a Majorana qubit. Let us investigate the Andreev qubit dynamics.

\subsection{Adiabatic evolution}

\begin{figure}[tb]
\centerline{\includegraphics[width=0.9\linewidth]{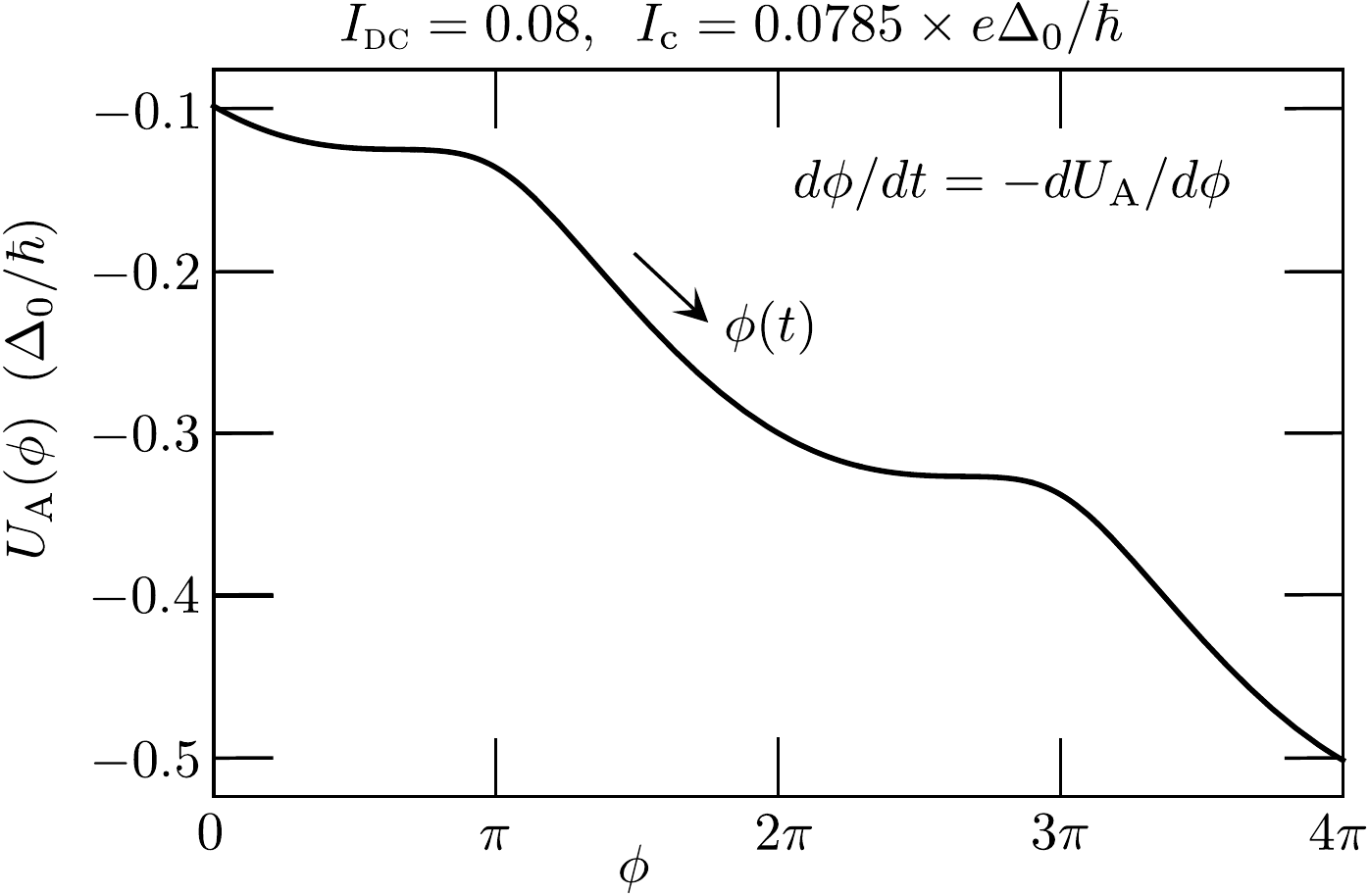}}
\caption{Washboard potential \eqref{washboardU} that governs the time dependence of the superconducting phase in the adiabatic limit. The curve is plotted for the junction parameters of Figs.\ \ \ref{fig_Evsphi} and \ref{fig_IV}a, at a value of $I_{\textsc{dc}}$ slightly above the critical current $I_{\rm c}$.
}
\label{fig_washboard}
\end{figure}

In the adiabatic regime of a slow driving, $\hbar\dot{\phi}\ll\delta E$, transitions between the Andreev levels can be neglected and the phase evolves in time as an overdamped classical particle,
\begin{equation}
\dot{\phi}+dU_{\rm A}/d\phi=0,\label{adiabaticeq}
\end{equation}
moving in the ``washboard potential'' \cite{tinkham}
\begin{equation}
U_{\rm A}(\phi)=-(2eR/\hbar)\bigl[\phi I_{\textsc{dc}}+(2e/\hbar)E_{\rm A}(\phi)\bigr],\label{washboardU}
\end{equation}
plotted in Fig.\ \ref{fig_washboard}.

\begin{figure*}[tb]
\centerline{\includegraphics[width=0.7\linewidth]{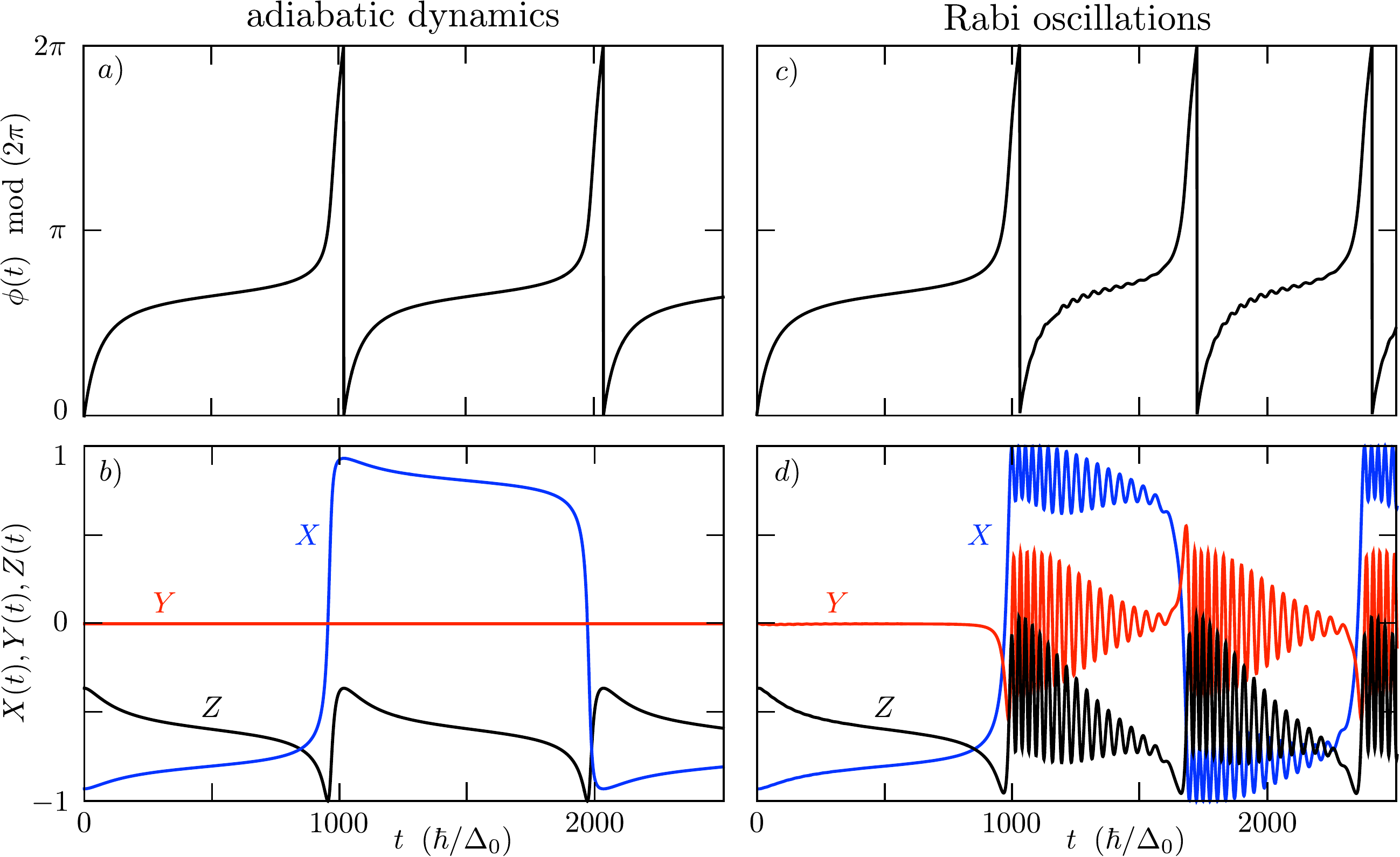}}
\caption{Time dependence of the superconducting phase (top row) and of the Bloch sphere coordinates of the Andreev qubit (bottom row), in the adiabatic limit (left column) and in the non-adiabatic regime in which transitions between the Andreev levels produce Rabi oscillations of the qubit (right column). The junction parameters are those of Fig.\ \ref{fig_IV}a, at $I_{\textsc{dc}}=0.08\,e\Delta_0/\hbar$. The wave function was initialized as an eigenstate of the lowest Andreev level $-E_{\rm A}(0)$ at $t=0$.
}
\label{fig_rabi}
\end{figure*}

The time dependence of the phase resulting from integration of Eq.\ \eqref{adiabaticeq} is shown in panel a) of Fig.\ \ref{fig_rabi}. Panel b) tracks the adiabatic dynamics of the Andreev qubit, by plotting the Bloch sphere coordinates  $\bm{R}=(X,Y,Z)$, with $R_\alpha(t)=\langle\Psi(t)|\sigma_\alpha|\Psi(t)\rangle$. The qubit dynamics is $4\pi$-periodic in $\phi$, because the Hamiltonian \eqref{Htdef} is $4\pi$-periodic: When $\phi$ is increased by $2\pi$ one has ${\cal H}\mapsto \sigma_z{\cal H}\sigma_z$, so on the Bloch sphere the qubit is rotated by $\pi$ around the $z$-axis ($X\mapsto -X$, $Y\mapsto-Y$). The full spectrum is a $2\pi$-periodic function of $\phi$, in particular the Josephson current \eqref{Ieqphi} is $2\pi$-periodic --- this nontopological Josephson junction does not exhibit the $4\pi$-periodic Josephson effect that is the hallmark of a topological superconductor.

\subsection{Pulsed Rabi oscillations}

Panels c) and d) of Fig.\ \ref{fig_rabi} show the full non-adiabatic dynamics, obtained by integration of Eq.\ \eqref{Schreq} for the same parameter set as in panels a) and b). Transitions between the Andreev levels produce pronounced Rabi oscillations of the qubit, also visible as small oscillations in $\phi(t)$. 

Because the supercurrent carried by the two Andreev levels $\pm E_{\rm A}$ has the opposite sign, the inter-level transitions reduce $I_{\rm S}$, thereby increasing $I_{\rm N}=I_{\textsc{dc}}-I_{\rm S}$ and hence $\bar{V}$. This is evident from Fig.\ \ref{fig_rabi}c, which shows that the first $2\pi$ increment of $\phi$, without interlevel transitions, takes a time $\delta t\approx 1000\,\hbar/\Delta_0$, while the second $2\pi$ increment, with Rabi oscillations, only takes a time $\delta t=700$. The average voltage $\bar{V}\simeq 2\pi/\delta t$ is therefore increased by a factor $10/7$ because of the interlevel transitions.

The Rabi oscillations are pulsed: they appear abruptly when $\phi$ crosses $(2n-1)\pi$ and increases rapidly to $2n\pi$, which is the steepest part of the washboard potential (see Fig.\ \ref{fig_washboard}). 

To estimate the Rabi frequency we substitute $\Psi(t)=\bigl(u(t)e^{i\phi(t)/4},v(t)e^{-i\phi(t)/4}\bigr)$ in the Schr\"{o}dinger equation \eqref{Schreq} and make the rotating wave approximation, discarding rapidly oscillating terms $\propto e^{i\phi(t)}$:
\begin{equation}
\begin{split}
&i\hbar \dot{u}(t)=[E_0-\mu+\tfrac{1}{2}eV(t)]u(t)+\tfrac{1}{4}\Gamma v(t),\\
&i\hbar \dot{v}(t)=-[E_0-\mu+\tfrac{1}{2}eV(t)]v(t)+\tfrac{1}{4}\Gamma u(t).
\end{split}
\end{equation}
(We have set $\Gamma_{\rm L}=\Gamma_{\rm R}$ for simplicity.) If we further neglect the slow time dependence of the voltage, we obtain oscillations $\propto \sin^2 \omega_{\rm R}t$ of the Bloch vector components $X,Y,Z$ with Rabi frequency
\begin{equation}
\hbar\omega_{\rm R}=\sqrt{(E_0-\mu+\tfrac{1}{2}eV)^2+(\Gamma/4)^2}.\label{eqRabi}
\end{equation}
The oscillations in Fig.\ \ref{fig_rabi}d near $t=1000\times \hbar/\Delta_0$ have a period of $35\,\hbar/\Delta_0$, while $T_{\rm R}=\pi/\omega_{\rm R}=40\,\hbar/\Delta_0$ if we set $V=RI_{\textsc{dc}}$, in reasonable agreement.

\subsection{Voltage steps count Rabi oscillations}
\label{sec_steps}

\begin{figure*}[tb]
\centerline{\includegraphics[width=0.7\linewidth]{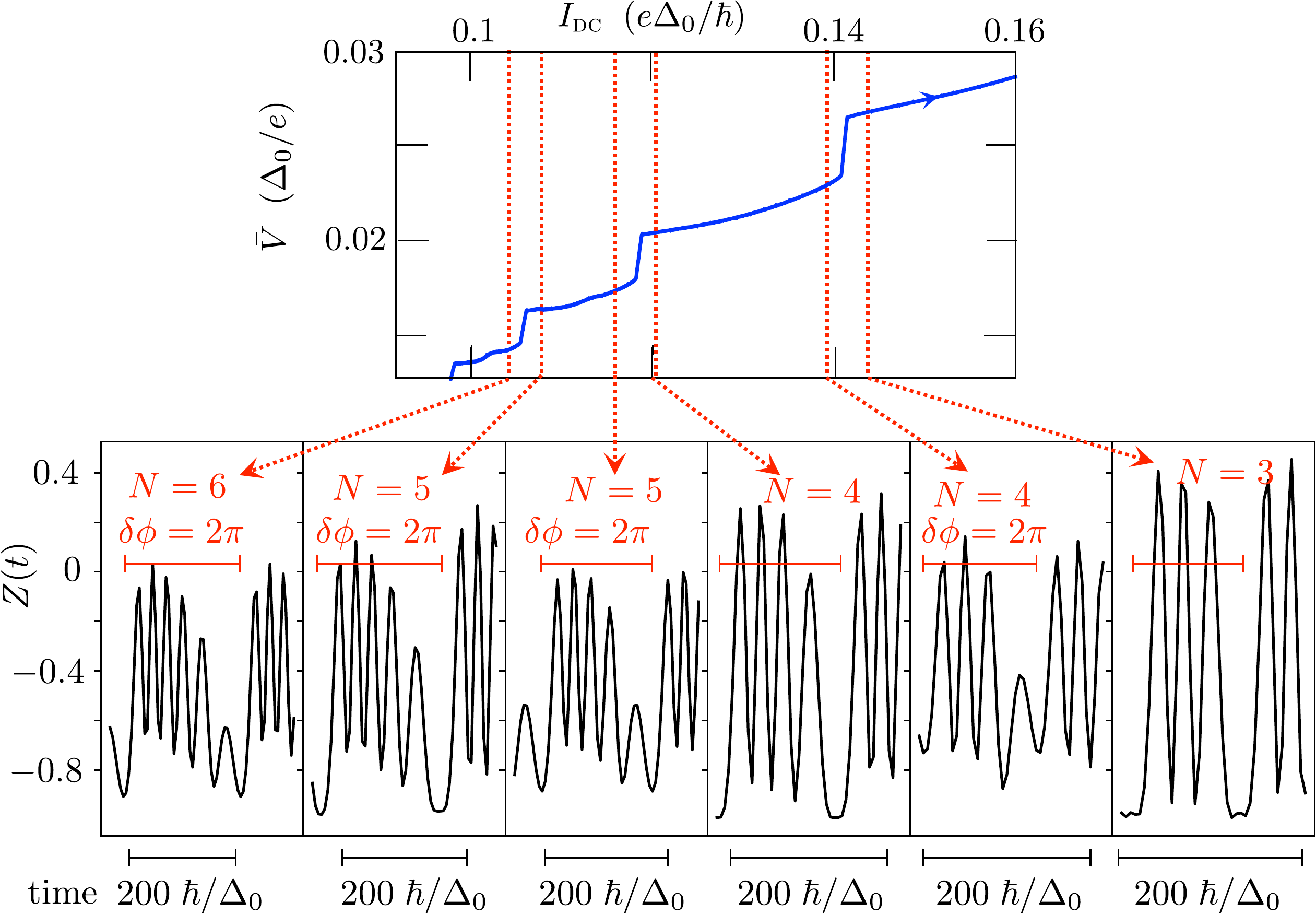}}
\caption{Top panel: portion of the $I$--$V$ characteristic from Fig.\ \ref{fig_IV}a, with red dotted lines into the the bottom panels to show how the voltage steps line up with the change in the number $N$ of Rabi oscillations of the qubit in a $2\pi$ phase increment $\delta\phi$.
}
\label{fig_counter}
\end{figure*}

The key discovery of Ref.\ \onlinecite{Cho20} is that steps in the time-averaged voltage  track the change in the number of Rabi oscillations of the Majorana qubit per $2\pi$ increment of the superconducting phase. Fig.\ \ref{fig_counter} shows the same correspondence for the Andreev qubit.

If we estimate the duration $\delta t$ of a $2\pi$ phase increment by the product of the number $N$ of Rabi oscillations and the Rabi period $T_{\rm R}$, we obtain the estimate $(2e/\hbar)\bar{V}=2\pi/\delta t\simeq 2\omega_{\rm R}/N$. A stepwise decrease of $N$ with increasing $I_{\textsc{dc}}$ would then produce a stepwise increase of $\bar{V}$. This argument is suggestive, but does not explain the sharpness of the steps. We have no quantitative analytical derivation for why the steps are as sharp as they appear in the numerics.

\section{Discussion}
\label{sec_discuss}

Two lessons learned from this study are: 1) Rabi oscillations of an Andreev qubit can be counted ``one-by-one'' without either requiring time-resolved detection or \textsc{ac} driving; 2) The voltage staircase phenomenology of Ref.\ \onlinecite{Cho20} does not need a topological Majorana qubit --- it exists in a conventional Andreev qubit.

We worked in the weak-coupling regime $\Gamma\ll\Delta_0$ because it simplifies the calculations, but also for a physics reason: The voltage staircase is suppressed when $\Gamma$ becomes larger than $\Delta_0$, due to a well-known decoherence mechanism \cite{Ave96,Ave98}: Equilibration of the Andreev levels $\pm E_{\rm A}(\phi)$ with the continuous spectrum at $|E|>\Delta_0$ when $\phi$ crosses an integer multiple of $2\pi$. Let us discuss this in a bit more detail.

For $\Gamma\gg\Delta_0$ the Andreev levels are given by
\begin{equation}
E_{\rm A}=\Delta_0\sqrt{1-\tau_{\rm BW}\sin^2(\phi/2)},
\end{equation}
according to Eq.\ \eqref{Omegaeq}, with $\tau_{\rm BW}$ the Breit-Wigner transmission probability \eqref{GBW}. The difference with the weak-coupling result \eqref{EAweakcoupling} is that the reduced gap $\Delta_{\rm eff}$ has been replaced by the true gap $\Delta_0$. This means the Andreev level merges with the superconducting continuum whenever $\phi=0$ modulo $2\pi$. As the phase evolves in time in response to the current bias, each $2\pi$ phase increment will restart from an equilibrium distribution.

Now if we examine Fig.\ \ref{fig_rabi}, panels c) and d), we see that the Rabi oscillations are pulsed by the rapid increase of the phase in the $(\pi,2\pi)$ interval, and only fully develop in the $(2\pi,3\pi)$ interval. Equilibration at $\phi=2\pi$ will restart the cycle from $t=0$, suppressing the Rabi oscillations and hence the voltage staircase.

For the same reason a superconducting quantum point contact will not show the voltage staircase: its Andreev levels also reconnect with the superconducting continuum at $\phi=0$ modulo $2\pi$.

This argument points to one difference in the Majorana versus Andreev phenomenology of the voltage staircase: A topological Josephson junction needs to be magnetic in order to prevent the equilibration of the Majorana modes with the continuum at $\phi=0$ modulo $2\pi$ \cite{Fu09}. In a non-topological quantum-dot Josephson junction this can achieved  without breaking time-reversal symmetry.

As a topic for further research, it would be worthwhile to see if the voltage staircase can be used to count the number of Rabi oscillations over multiple $2\pi$ phase increments, since that would provide additional information on the coherence time of the qubit. This could involve the constructive interference of Landau-Zener transitions at $\phi=\pi,3\pi,\ldots$ \cite{She10}.
 
\acknowledgments
We have benefited from discussions with A. R. Akhmerov, F. Hassler, V. S. Shumeiko, T. Vakhtel, B. van Heck, and V. Verteletskyi. This project has received funding from the Netherlands Organization for Scientific Research (NWO/OCW) and from the European Research Council (ERC) under the European Union's Horizon 2020 research and innovation programme.

\appendix

\section{Hysteresis of the voltage staircase for the Majorana qubit}
\label{app_hysteresis}

The voltage staircase of the Andreev qubit is hysteretic, the steps appear at higher current for the up-sweep than for the down-sweep. No hysteresis was reported in Ref.\ \onlinecite{Cho20}, here we show that it is present for the Majorana qubit as well.

Instead of Eqs.\ \eqref{Htdef} and \eqref{Itdef} one has for the Majorana qubit the time dependent Hamiltonian
 \begin{equation}
{\cal H}(t)=E_x\sigma_x+E_{z}\sigma_z\cos\tfrac{1}{2}\phi(t),\label{HMajorana}
\end{equation}
and current operator
\begin{equation}
{I}(t)=\frac{2e}{\hbar}\frac{\partial}{\partial \phi}{\cal H}(t)=-\frac{eE_{z}}{\hbar}\sigma_z\sin\tfrac{1}{2}\phi(t).
\end{equation}

The Pauli matrices act on the fermion parity of two pairs of Majorana zero-modes, such that $\sigma_x$ flips the even--even parity state into the odd--odd parity state, while $\sigma_z$ changes the sign of the odd--odd parity state. While the physical origin of the Majorana coupling terms is different from the Andreev qubit, mathematically the Hamiltonian \eqref{HMajorana} is equivalent to Eq.\ \eqref{Htdef} in the symmetric case $\Gamma_{\rm L}=\Gamma_{\rm R}$. (Switch $\sigma_x\leftrightarrow\sigma_z$ by a unitary transformation and replace $E_x\mapsto E_0-\mu$ and $E_z\mapsto \Gamma/2$.)

\begin{figure}[tb]
\centerline{\includegraphics[width=0.9\linewidth]{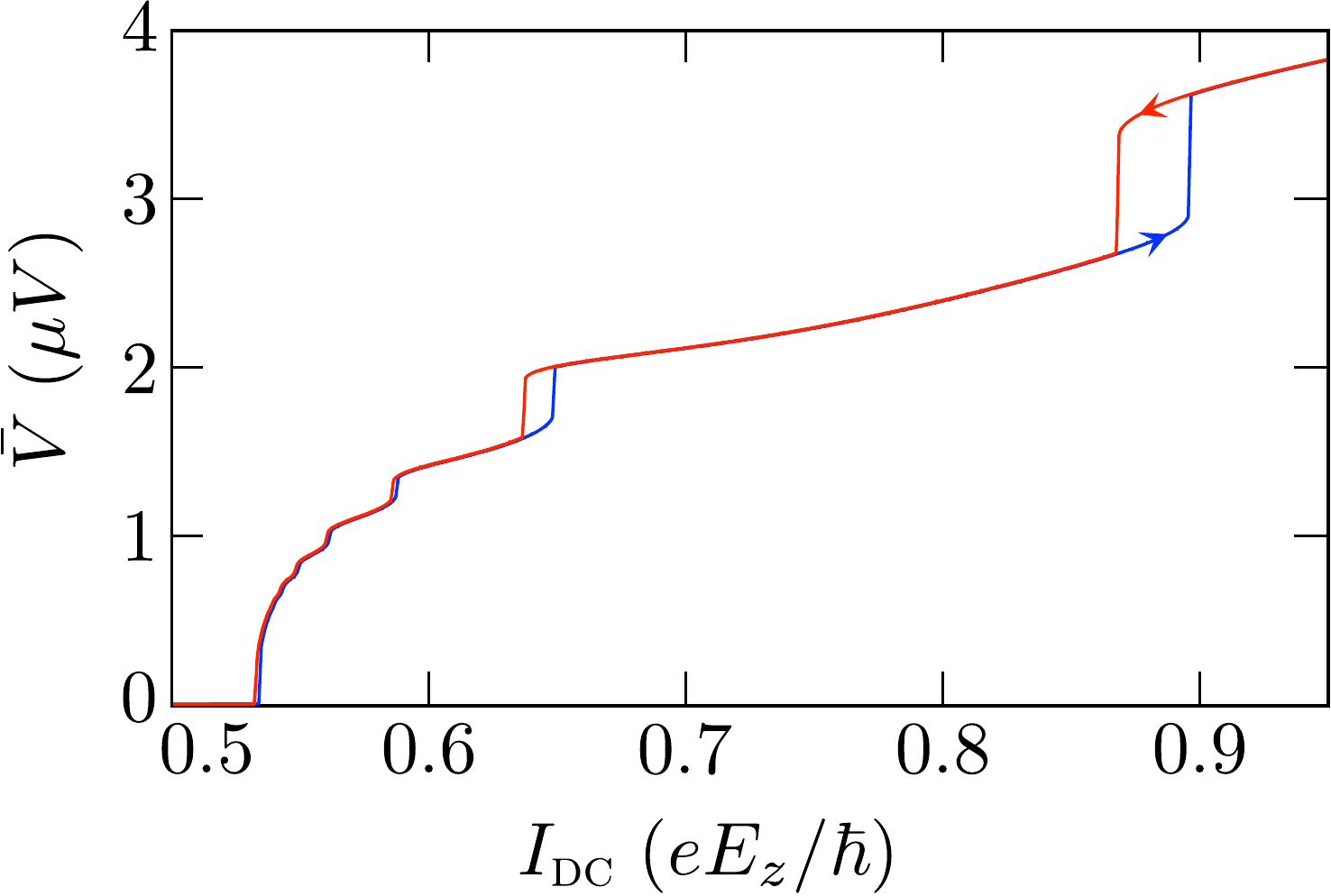}}
\caption{Hysteretic voltage staircase of the Majorana Josephson junction, for the parameters of Ref.\ \onlinecite{Cho20}, Fig.\ 3. The blue curve is for increasing {\sc dc} current, the red curve for decreasing current. (The voltage $\bar{V}$ is averaged over a time window $\delta t$ such that $\delta t\times dI_{\textsc{dc}}/dt= 10^{-3}\,eE_z/\hbar$.)
}
\label{fig_IVnew}
\end{figure}

In Fig.\ \ref{fig_IVnew} we show the hysteretic voltage staircase, for the same parameters $E_z=5\,\mu\text{eV}$, $E_x/E_{z}=0.67$, $R=0.827\,\hbar/e^2$ as in Ref.\ \onlinecite{Cho20}.

\begin{thebibliography}{99}
\bibitem{note1} The time-dependent unitary transformation $\Psi\mapsto U^\dagger\Psi$, ${\cal H}\mapsto U^\dagger{\cal H}U-i\hbar U^\dagger dU/dt$ with $U(t)=e^{i\sigma_z\gamma\phi(t)/4}$ removes the $\dot{\phi}$-term from the Hamiltonian \eqref{Htdef}. The $\gamma$-parameter then appears in the superconducting phases, $\phi_{\rm L}=\frac{1}{2}(1-\gamma)\phi$, $\phi_{\rm R}=-\frac{1}{2}(1+\gamma)\phi$.
\bibitem{note2} The parameters $E_0,\Gamma_{\rm L},\Gamma_{\rm R}$ used in Fig.\ \ref{fig_IV} are listed in each panel; additional parameters: $\mu=0$ in both panels, $R=0.20$ and $0.25\,\hbar/e^2$ in panels a) and b), respectively. The voltage $\bar{V}$ is averaged over a time window $\delta t$ such that $\delta t \times dI_{\textsc{dc}}/dt=6.3\cdot 10^{-4}\,e\Delta_0/\hbar$. To check that we are sweeping slowly enough, we reduced $dI_{\textsc{dc}}/dt$ by a factor of two and found little difference.
\end{thebibliography}
\end{document}